\documentstyle[12pt]{article}

\catcode`@=11
\def\un#1{\relax\ifmmode\@@underline#1\else
        $\@@underline{\hbox{#1}}$\relax\fi}
\catcode`@=12

\def\a{\alpha}
\def\b{\beta}
\def\c{\chi}
\def\d{\delta}
\def\e{\epsilon}
\def\f{\phi}
\def\g{\gamma}
\def\h{\eta}

\def\j{\psi}

\def\l{\lambda}
\def\m{\mu}

\def\o{\omega}
\def\p{\pi}
\def\q{\theta}

\def\x{\xi}
\def\z{\zeta}
\def\D{\Delta}

\def\O{\Omega}

\def\S{\Sigma}

\def\ve{\varepsilon}
\def\vf{\varphi}

\def\vq{\vartheta}

\def\cm{{\cal M}}
\def\cn{{\cal N}}

\def\bo{{\raise-.5ex\hbox{\large$\Box$}}}               
\def\pa{\partial}                                       
\def\de{\nabla}                                         
\def\TH{{\raise.2ex\hbox{$\displaystyle \bigodot$}\mskip-4.7mu \llap H \;}}
\def\face{{\raise.2ex\hbox{$\displaystyle \bigodot$}\mskip-2.2mu \llap {$\ddot
        \smile$}}}                                      
\def\dg{\sp\dagger}                                     

\def\sp#1{{}^{#1}}                              
\def\VEV#1{\left\langle #1\right\rangle}        
\def\abs#1{\left| #1\right|}                    
\def\leftrightarrowfill{$\mathsurround=0pt \mathord\leftarrow \mkern-6mu
        \cleaders\hbox{$\mkern-2mu \mathord- \mkern-2mu$}\hfill
        \mkern-6mu \mathord\rightarrow$}
\def\dvec#1{\vbox{\ialign{##\crcr
        \leftrightarrowfill\crcr\noalign{\kern-1pt\nointerlineskip}
        $\hfil\displaystyle{#1}\hfil$\crcr}}}           

\def\frac#1#2{{\textstyle{#1\over\vphantom2\smash{\raise.20ex
        \hbox{$\scriptstyle{#2}$}}}}}                   
\def\sfrac#1#2{{\vphantom1\smash{\lower.5ex\hbox{\small$#1$}}\over
        \vphantom1\smash{\raise.4ex\hbox{\small$#2$}}}} 
\def\bfrac#1#2{{\vphantom1\smash{\lower.5ex\hbox{$#1$}}\over
        \vphantom1\smash{\raise.3ex\hbox{$#2$}}}}       
\def\afrac#1#2{{\vphantom1\smash{\lower.5ex\hbox{$#1$}}\over#2}}    

\def\[{\lfloor{\hskip 0.35pt}\!\!\!\lceil}
\def\]{\rfloor{\hskip 0.35pt}\!\!\!\rceil}

\def\un{\underline}
\def\fracmm#1#2{{{#1}\over{#2}}}

\def\low#1{{\raise -3pt\hbox{${\hskip 0.75pt}\!_{#1}$}}}

\newskip\humongous \humongous=0pt plus 1000pt minus 1000pt
\def\caja{\mathsurround=0pt}
\def\eqalign#1{\,\vcenter{\openup2\jot \caja
        \ialign{\strut \hfil$\displaystyle{##}$&$
        \displaystyle{{}##}$\hfil\crcr#1\crcr}}\,}
\newif\ifdtup

\def\ref#1{$\sp{#1)}$}

\def\pl#1#2#3{Phys.~Lett.~{\bf {#1}B} (19{#2}) #3}
\def\np#1#2#3{Nucl.~Phys.~{\bf B{#1}} (19{#2}) #3}

\def\cmp#1#2#3{Commun.~Math.~Phys.~{\bf {#1}} (19{#2}) #3}

\def\mpl#1#2#3{Mod.~Phys.~Lett.~{\bf A{#1}} (19{#2}) #3}

\parskip=\medskipamount                 
\thicklines                         

\begin{document}


\baselineskip=20pt
\def\sfo{S\!F\!O}



\thispagestyle{empty}               

\def\border{                                            
        \setlength{\unitlength}{1mm}
        \newcount\xco
        \newcount\yco
        \xco=-24
        \yco=12
        \begin{picture}(140,0)
        \put(-20,11){\tiny Institut f\"ur Theoretische Physik Universit\"at
Hannover~~ Institut f\"ur Theoretische Physik Universit\"at Hannover~~
Institut f\"ur Theoretische Physik Hannover}
        \put(-20,-241.5){\tiny Institut f\"ur Theoretische Physik Universit\"at
Hannover~~ Institut f\"ur Theoretische Physik Universit\"at Hannover~~
Institut f\"ur Theoretische Physik Hannover}
        \end{picture}
        \par\vskip-8mm}

\def\headpic{                                           
        \indent
        \setlength{\unitlength}{.8mm}
        \thinlines
        \par
        \begin{picture}(29,16)
        \put(75,16){\line(1,0){4}}
        \put(80,16){\line(1,0){4}}
        \put(85,16){\line(1,0){4}}
        \put(92,16){\line(1,0){4}}

        \put(85,0){\line(1,0){4}}
        \put(89,8){\line(1,0){3}}
        \put(92,0){\line(1,0){4}}

        \put(85,0){\line(0,1){16}}
        \put(96,0){\line(0,1){16}}
        \put(79,0){\line(0,1){16}}
        \put(80,0){\line(0,1){16}}
        \put(89,0){\line(0,1){16}}
        \put(92,0){\line(0,1){16}}
        \put(79,16){\oval(8,32)[bl]}
        \put(80,16){\oval(8,32)[br]}

        \end{picture}
        \par\vskip-6.5mm
        \thicklines}

\border\headpic {\hbox to\hsize{
\vbox{\noindent ITP--UH--13/95 \hfill March 1995 \\
hep-th/9503232 \hfill }}}

\noindent
\vskip1.3cm
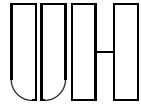
\begin{center}

{\Large\bf The String Measure and Spectral Flow \\
		of Critical $~N=2~$ Strings}
\footnote{Supported in part by the `Deutsche Forschungsgemeinschaft'}\\
\vglue.3in

Sergei V. Ketov \footnote{
On leave of absence from:
High Current Electronics Institute of the Russian Academy of Sciences,
\newline ${~~~~~}$ Siberian Branch, Akademichesky~4, Tomsk 634055, Russia}
and  Olaf Lechtenfeld \vglue.1in
{\it Institut f\"ur Theoretische Physik, Universit\"at Hannover}\\
{\it Appelstra\ss{}e 2, 30167 Hannover, Germany}\\
{\sl ketov, lechtenf @itp.uni-hannover.de}
\end{center}

\vglue.2in
\begin{center}
{\Large\bf Abstract}
\end{center}

The general structure of $N{=}2$ moduli space at arbitrary genus {\it and\/}
instanton number is investigated. The $N{=}2$ NSR string measure is calculated,
yielding picture- and $U(1)$ ghost number-changing operator insertions.
An explicit formula for the spectral flow operator acting on vertex operators
is given, and its effect on $N{=}2$ string amplitudes is discussed.

\newpage

{\bf 1} {\it Introduction}. The recent interest \cite{bv,b,o} in computing
loop scattering amplitudes of critical $N{=}2$ strings in $2{+}2$ dimensions
is partially motivated by a rich structure of $N{=}2$ moduli space, which leads
to some novel features of string perturbation theory when compared to the
$N{=}1$ and $N{=}0$ cases. In refs.~\cite{bv,b,o}, an $N{=}4$ (twisted)
supersymmetrical topological description of critical $N{=}2$ strings was
used to bypass problems with the (bosonized) $N{=}2$ superconformal ghosts and
the location of the $N{=}2$ picture-changing operators. Still, it is of
interest
to pursue the non-topological approach of integrating correlation functions
of BRST-invariant vertex operators over all $N{=}2$ supergravity moduli.
We shall show, in particular, how the integration over $U(1)$ moduli
generates $U(1)$ ghost number-changing operators
and implements the spectral flow for string amplitudes.

In refs.~\cite{klp,bkl,l}, we have investigated the critical $N{=}2$ string in
the BRST formalism, but confined ourselves to calculating the BRST cohomology
and tree amplitudes without $U(1)$ instantons. In this letter, we are going to
describe the structure of the $N{=}2$ moduli space at arbitrary genus
{\it and\/} $U(1)$ instanton number, and calculate the string measure resulting
from integrating out fermionic as well as $U(1)$ moduli.

The gauge-invariant $N{=}2$ string world-sheet action is given by the minimal
coupling of $N{=}2$ matter represented by two $N{=}2$ scalar multiplets
$(X^\m,\j^\m)$ \footnote{
The fields $X^\m$ and $\j^\m$ are complex-valued, and $\m=0,1$ is a vector
index
w.r.t. the spacetime `Lorentz group' $U(1,1)$.}
to $N{=}2$ supergravity comprising a
zweibein $e^a_{\a}$ or a metric $g_{\a\b}$, a real $U(1)$ gauge field $A_{\a}$,
and two real gravitini $\c^{\pm}_{\a}$, \footnote{
The superscript $\pm$ of a field denotes its $U(1)$ charge $\pm 1$, and
$\a,\b=0,1$ are 2d world-sheet indices.}
all living on the two-dimensional (2d) world-sheet.
The local symmetries are given
by 2d reparametrizations, local Lorentz and Weyl invariances,
$N{=}2$ supersymmetry and $N{=}2$ super-Weyl symmetry,
phase $U_{\rm V}(1)$ and chiral $U_{\rm A}(1)$ invariances. \footnote{
Note that there is only a single gauge field for both $U(1)$ invariances.}
The world-sheet $\S$ is supposed to arise from a closed orientable $N{=}2$
super-Riemann surface, whose topology is characterized by two integers,
its Euler number $\c$ (related to the genus $h$) and its first Chern class
(or $U(1)$ instanton number \footnote{
After Wick-rotating the world-sheet to the Euclidean regime, it is better
to think of $c$ as a $U(1)$ monopole charge.}
) $c\,$,
$$
\c=\fracmm{1}{2\p}\int_{\S}R~=2-2h~,\quad h\in {\bf N}~;\qquad
c=\fracmm{1}{2\p}\int_{\S}F~=k~,\quad k\in {\bf Z}~.
\eqno(1)$$
Here, the curvature two-form $R$ and the $U(1)$ field strength two-form $F=dA$
have been introduced.
Finally, correlation functions require the introduction of a third integer,
$n\in{\bf N}$, which counts the number of punctures on~$\S_{h,n}$.
\vglue.1in

{\bf 2} {\it The U(1) bundles}.
The $U(1)$ gauge field $A_{\a}=(A_z,A_{\bar{z}})$ is the vertical connection
on a principal $U(1)$ bundle over $\S_{h,n}$.
It can always be split as $A=A^{\rm inst} + A^0$, where the
instantonic part $A^{\rm inst}$ saturates the instanton number,
$c(A^{\rm inst})=k$, and $A^0$ is globally defined on $\S_{h,n}$ since
$c(A^0)=0$.
The ambiguity in choosing a representative $A^{\rm inst}$ is contained
in the space of~$A^0$'s.
Since we are going to integrate over the moduli space of all $\,U(1)$
connections, we need to decompose $A^0$ into gauge and moduli parts.
The Hodge decomposition on $\S_{h,n}$ reads
$$
A^0 \ =\ d\f +\d \o + A^{\rm Teich}~,
\eqno(2)$$
where the function $\f$ and two-form $\o$ represent the $U_{\rm V}(1)$ and
$U_{\rm A}(1)$ gauge degrees of freedom, respectively,
and $A^{\rm Teich}$ is a harmonic one-form on $\S_{h,n}$
representing the $U(1)$ Teichm\"uller degrees of freedom. \footnote{
We use the notation $\d=*d*$, where the star means Hodge conjugation,
so that $\D=\d d+d\d$ and $d^2=0=\d^2$.
The connection $A^{\rm Teich}$ is both closed and co-closed,
$dA^{\rm Teich}=0=\d A^{\rm Teich}$.}
In the presence of punctures, however, this decomposition is not yet unique.
There exist harmonic one-forms which are also exact or co-exact but diverge
at the punctures. In locally flat holomorphic coordinates around a puncture,
the co-exact prototype is
$$
\tilde A\ =\ {1\over2i}\left({dz\over z}-{d\bar z\over\bar z}\right)
\ =\ d\tilde\f\ =\ \d\tilde\o \quad,\qquad
\tilde\f\ =\ {1\over2i}\ln {z\over\bar z} \quad,\quad
\tilde\o\ =\ {i\over2}\ln|z|\,dz\wedge d\bar z
\eqno(3)$$
which is harmonic because
$$
\tilde F\ =\ d\tilde A\ =\ 2\p i\,\d^{(2)}(z)\,dz\wedge d\bar z
\eqno(4)$$
vanishes away from the puncture, implying $\D\tilde A=0$ at least locally
on $\S_{h,n}$.
$\tilde A$ is not exact because $\tilde\f$ is multi-valued around $z{=}0$.
Clearly, $\l\tilde A$ represents a Dirac monopole with magnetic charge~$\l$.
\footnote{
Globally, the condition $c(A^0)=0$ forces the sum of all magnetic charges
to vanish.}
Similarly, one may utilize the Hodge dual of $\tilde A$,
$$
*\tilde A\ =\ -{1\over2}\left({dz\over z}+{d\bar z\over\bar z}\right)
\ =\ \d *\tilde\f\ =\ -d *\tilde\o \quad,
\eqno(5)$$
to manufacture an exact (but not co-exact) harmonic one-form
which does not contribute to the curvature~$F$ at all.
We shall render the decomposition (2) unique by putting such gauge fields
into the $A^{\rm Teich}$ part, since they do not really represent
$U_{\rm V}(1)$ or $U_{\rm A}(1)$ gauge degrees of freedom but do something
non-trivial to the punctures.
With $\f$ and $\o$ being smooth at the punctures, the $U_{\rm A}(1)$ gauge
symmetry changes $F$ but not $c$.
Likewise, the $\,U(1)$ Teichm\"uller variations must respect $c(A^0)=0$,
$$
0\ =\ \fracmm{1}{2\p}\int_{\S_{h,n}}dA^0\ =\
-\fracmm{1}{2\p}\sum_{\ell=1}^n \oint_{z_\ell} A^{\rm Teich} \quad.
\eqno(6)$$
A second condition arises from the fact that the Lorentz gauge condition,
$\d A=0$, can be chosen globally,
$$
0\ =\ \fracmm{1}{2\p}\int_{\S_{h,n}}*\d A^0\ =\
-\fracmm{1}{2\p}\sum_{\ell=1}^n \oint_{z_\ell} *A^{\rm Teich} \quad.
\eqno(7)$$

The complex structure of $\S$ naturally suggests the complex linear
combinations
$$
A^\pm\ =\ A^0\pm i*A^0 \quad,\qquad d^\pm\ =\ d\pm i*d \quad,\qquad
\f^\pm\ =\ \f\mp i*\o \quad
\eqno(8)$$
so that
$$
A^\pm\ =\ d^\pm\f^\pm+A_{\rm Teich}^\pm \quad,\qquad
F^\pm\ =\ dA^\pm\ =\ F^0\pm i*\d A^0 \quad,
\eqno(9)$$
and the singular one-forms of eqs. (3) and (5) combine to
$$
\tilde A^+\ =\ {dz\over z}\quad,\qquad\tilde A^-\ =\ {d\bar z\over\bar z}\quad.
\eqno(10)$$

Independent of $c$,
it is easy to count the dimension of the $U(1)$ Teichm\"uller space.
On $\S_{h,n}$, one has $2h$ real abelian one-forms $\a_i$ and $\b_i$,
$i=1,\dots,h$, which may be chosen to be dual to the basis $\{a_i,b_i\}$ of
homology cycles,
$$
\oint_{a_i}\a_j=\d_{ij}\quad,\qquad \oint_{b_i}\a_j=0\quad,\qquad
\oint_{a_i}\b_j=0\quad,\qquad \oint_{b_i}\b_j=\d_{ij}\quad,
\eqno(11)$$
as well as $2n$ real one-forms of the type given in eqs. (3)
and (5), namely two for each puncture.
However, eqs. (6) and~(7) put two real constraints on the
coefficients multiplying the latter,
so that the total real dimension equals $2h+2n-2$ for $n{>}0$.

Not all $U(1)$ Teichm\"uller variations are moduli, however.
Whenever the Wilson loops
$$
W[A;\g]\ =\ \exp \Bigl\{ i\oint_\g A \Bigr\} \qquad {\rm and} \qquad
*\!W[A;\g]\ =\ \exp \Bigl\{ i\oint_\g *A \Bigr\}
\eqno(12)$$
become trivial ($=1$) for some $A^{\rm Teich}$
and all cycles $\g$, we have encountered a `big' $U(1)$ gauge transformation.
Invariably, this will happen if the coefficients multiplying the harmonic
one-forms get large enough, as for $\l=1$ in
$$
1 = \exp\Bigl\{i\oint_0\l\tilde A\Bigr\} = e^{2\p i\l} \qquad
\Longleftrightarrow\qquad
i\tilde A = g^{-1} dg \quad{\rm with}\quad g(z)=e^{i\tilde\f(z)}={z\over|z|}~,
\eqno(13)$$
or for $\vf=1$ in
$$
1 = \exp\Bigl\{i\oint_{a_1} 2\p\vf\a_1\Bigr\} = e^{2\p i\vf} \qquad
\Longleftrightarrow\qquad
2\p i\a_1 = g^{-1} dg \quad{\rm with}\quad g(z)=e^{2\p i\int^z \a_1}~.
\eqno(14)$$
Therefore, the $U(1)$ moduli space of $\S_{h,n}$ is parametrized
by $2h+2n-2$ real phases (twists),
$$
\oint_{a_i}A = 2\p\vf_i \quad,\qquad
\oint_{b_i}A = 2\p\vq_i \quad,\qquad
\oint_{z_\ell}A = 2\p\l_\ell \quad,\qquad
\oint_{z_\ell}*A= 2\p\m_\ell \quad,
\eqno(15)$$
subject to $\sum_\ell \l_\ell=0=\sum_\ell \m_\ell$.
All phases range from $0$ to $1$.
It is important to note, however, that one may change the value of~$c$
simply by shifting the $\l_\ell$ so that the total phase shift is integral.
In summary, one obtains (for $n{>}0$) a real torus
$$
\cm^{U(1)}_{h,n}\ =\ {{\bf R}^{2(h+n-1)}\over{\bf Z}^{2(h+n-1)}}~.
\eqno(16)$$
This picture does not depend on the value $k$ of the instanton number~$c$.
For a thorough discussion of such matters in the context of $N{=}0$ and
$N{=}1$ strings, see ref.~\cite{dp}.

{} From now on we shall employ the $U_{\rm V}(1)$ Lorentz gauge, $\d A^0=0$,
and its $U_{\rm A}(1)$ counterpart, $F^0=0$, on $\S_{h,n}$,
effectively setting $\f=\o=0$ in eq.~(2).
The $U(1)$ moduli space can then be seen as the moduli space of flat
connections on $\S_{h,n}$, which is a product of two factors.
One factor is the Jacobian variety of flat $U(1)$ connections on $\S_{h,0}$,
$$
J(\S_{h,0})\ =\ \fracmm{{\bf C}^h}{{\bf Z}^h+\O{\bf Z}^h}
\ \sim\ \fracmm{{\bf R}^{2h}}{{\bf Z}^{2h}}\ =\ Pic(h,0) \quad,
\eqno(17)$$
where $\O$ is the period matrix of $\S$.
It is diffeomorphic to the torus $Pic(h,0)$
para\-metrizing all flat holomorphic line bundles over $\S_{h,0}$, with
twists $\vf_i$ and $\vq_i$ of eq.~(15) on the homology cycles.
The other factor is the torus
$$
Pic(0,n)\ =\ \fracmm{{\bf R}^{2(n-1)}}{{\bf Z}^{2(n-1)}}
\eqno(18)$$
encoding the $2n-2$ independent twists $\l_\ell$ and $\m_\ell$ around the
punctures on the Riemann sphere~$\S_{0,n}$.
The corresponding flat connections are given by linear combinations of
the singular one-forms given in eq.~(10), for each puncture.

To construct the $N{=}2$ string measure, we need to know the instanton
solution $A^{\rm inst}$ and the flat connection $A^{\rm Teich}$ explicitly, in
terms of moduli. As far as $A^{\rm inst}$ is concerned, it can be chosen to
satisfy the Laplace-Beltrami equation (of motion) and the gauge condition
$\d A^{\rm inst}=0$ on the punctureless Riemann surface~$\S\equiv\S_{h,0}$.
Taken together with the conformal gauge for the 2d metric $g$,
they lead to the simple equations
$$
\pa_{\bar{z}}g^{z\bar{z}}\pa_zA^{\rm inst}_{\bar{z}}\  =\ 0\qquad {\rm and}
\qquad \pa_zA^{\rm inst}_{\bar{z}}+\pa_{\bar{z}}A^{\rm inst}_z\ =\ 0\quad,
\eqno(19)$$
respectively. Since $\S$ is compact, orientable and without boundary,
eq.~(19) implies that
$\pa_zA^{\rm inst}_{\bar{z}}=-\pa_{\bar{z}}A^{\rm inst}_z\sim g_{z\bar{z}}$.
The coefficient is easily fixed by the instanton number constraint,
$c(A^{\rm inst})=k$,
$$
\pa_zA^{\rm inst}_{\bar{z}}\ =\ -\pa_{\bar{z}}A^{\rm inst}_z\ =\
\fracmm{\p k}{A}g_{z\bar{z}}\quad,
\eqno(20)$$
where the total area $A=\int_\S d^2z\,g_{z\bar{z}}$
of $\S$ has been introduced.
The solution to eq.~(20) is  ({\it cf}. ref.~\cite{f})
$$\eqalign{
A^{\rm inst}_z(z,\bar{z})\ =&\ -\fracmm{\p k}{A}\int_{\S}d^2w\,\pa_z K(z,w)
g_{w\bar{w}}  +\p k \sum^h_{i,j=1}\o_i(z)\left|{\rm Im}\O_{ij}\right|^{-1}
\int_{\bar{z}_0}^{\bar{z}}\bar{\o}_j(\bar{w})d\bar{w}\quad,\cr
A^{\rm inst}_{\bar{z}}(z,\bar{z})\ =&\ +\fracmm{\p k}{A}\int_{\S}d^2w\,
\pa_{\bar{z}} K(z,w) g_{w\bar{w}}  - \p k \sum^h_{i,j=1}\bar{\o}_i(\bar{z})
\left|{\rm Im}\O_{ij}\right|^{-1} \int_{z_0}^z \o_j(w)dw\quad,\cr}
\eqno(21)$$
where $K(z,w)$ is the Green function to the scalar Laplacian,
$$
\pa_z\pa_{\bar{z}}K(z,w)\ =\ \d^{(2)}_{z\bar{z}}(z,w) +
\o_i(z)\left|{\rm Im}\O_{ij}\right|^{-1}\bar{\o}_j(\bar{z})\quad~,
\eqno(22)$$
and $\o_i=\a_i+\O_{ij}\b_j$, $i=1,\ldots,h$, are the holomorphic abelian
differentials with normalization
$$
\oint_{a_i}\o_j\ =\ \d_{ij}\qquad,\qquad \oint_{b_i}\o_j\ =\ \O_{ij}\quad.
\eqno(23)$$
The solution to eq.~(22) can be expressed in terms of the prime form $E(z,w)$
\footnote{
The prime form $E(z,w)$ is a holomorphic $(-\fracmm12,0)$ form in $z$ and
$w$, with a single zero at $z=w$.} as
$$
K(z,w)\ =\ \ln\left|E(z,w)\right|^2 + \sum^h_{i,j=1}\left[ {\rm Im}
\int^z_w \o_i(u)du\right] \left|{\rm Im}\O_{ij}\right|^{-1}
\left[ {\rm Im} \int^z_w \o_j(u)du\right]~.
\eqno(24)$$
The fields $A^{\rm inst}_{z,\bar{z}}$ are neither holomorphic nor
single-valued around the homology cycles $a_i$ or $b_j$
but change by a gauge transformation (as long as $k$ is integer).

Turning to $A^{\rm Teich}$,
a convenient parametrization is given by
$$
A^{\rm Teich}\ =\ 2\p \sum_{i=1}^h (\vf_i\a_i + \vq_i\b_i) +
\sum_{\ell=1}^n (\l^+_\ell \tilde A^+_\ell + \l^-_\ell \tilde A^-_\ell)\quad,
\eqno(25)$$
where, in locally flat complex coordinates,
$$
A^+_\ell(z)\ =\ {dz\over z-z_\ell} \qquad{\rm and}\qquad
A^-_\ell(z)\ =\ {d\bar z\over \bar z-\bar z_\ell} \quad.
\eqno(26)$$
\vglue.1in

{\bf 3} {\it The gravitini bundles}.
The 2d gravitini $\c^{\pm}_{\a}$ transform inhomogeneously
under the $N{=}2$ local supersymmetry and $N{=}2$ super-Weyl (fermionic) gauge
symmetry as
$$
\d_{\rm S}\c^{\pm}_{\a}\ =\ \hat{D}_{\a}\ve^{\pm}\quad,\qquad
\d_{\rm W}\c^{\pm}_{\a}\ =\ \g_{\a}\z^{\pm}\quad,
\eqno(27)$$
where $\hat{D}_{\a}(\hat{\o},A)$ is the $N{=}2$ supergravitational covariant
derivative containing the spin connection $\hat{\o}_{\a}$ and the $U(1)$ gauge
connection $A_{\a}$.
The $\c^{\pm}_{\a}$ transform homogeneously under all other local symmetries,
with a $U(1)$ charge of~$\pm1$.
Hence, the gravitini are sections of some complex spinor bundle associated to
the principal $U(1)$ bundle over $\S_{h,n}$.

The local symmetries of eq.~(27) allow one to gauge away all 8 real
Grassmann degrees of freedom of $\c^{\pm}_{\a}$, except for those
in the kernel of $\hat{D}^{\pm}$,
$$
\hat{D}^{\pm} \c^\pm_z \ \equiv\
\left( \bar{\pa} \mp iA_{\bar{z}} \right) \c^\pm_z \ =\ 0 \quad,
\eqno(28)$$
and similarly for $\c^{\pm}_{\bar{z}}$ (signs are correlated).
We have used the superconformal gauge, in which $\hat{\o}_{\bar z}=0$
and all $\c$'s are $\g$-traceless.
The solutions of eq.~(28) are the fermionic moduli on $\S_{h,n}$.
Their number depends on $h$ and $n$ as well as on the instanton number~$k$
and is dictated by the Riemann-Roch theorem,
$$
{\rm ind} \hat{D}^{\pm}\ \equiv\
{\rm dim~ker}\,\hat{D}^{\pm} - {\rm dim~ker}\,\hat{D}^{\pm\dg}\
=\ 2(h-1)\pm k +n ~\quad.
\eqno(29)$$
For $h>1$, the contributions to a positive index generically come from the
first term, so that on reads off $2(h-1)+n+k$ positively charged and
$2(h-1)+n-k$ negatively charged fermionic moduli.
When the index becomes negative, $\det\hat{D}^{\pm\dg}$ develops
zero modes, which implies the vanishing of the corresponding $n$-point
correlator. This restricts the range of the sum over~$k$ to
$$
\abs{k}\ \leq\ 2(h-1)+n \quad.
\eqno(30)$$

The issue of a complex structure for the gravitini bundles with $k\ne0$
is a subtle one since $A_{\bar{z}}$ in $\hat{D}^{\pm}$ contains the
non-holomorphic $A_{\bar{z}}^{\rm inst}$ of eq.~(21).
Even without punctures ($n{=}0$), the gravitini bundles cannot in general
be {\it holomorphic\/} line bundles.
The latter are (twisted) integral or half-integral powers of the
canonical line bundle and as such always yield integral multiples of $h{-}1$
for ${\rm ind}\hat{D}$.
The r.h.s. of eq.~(29), however, is of this form only when $k$ itself is
a multiple of $h{-}1$, in which case the gauge connection may be absorbed
in the spin connection, effectively shifting the conformal weights of the
gravitini from $\fracmm32$ to $\fracmm32$, to $\fracmm32{\pm}\fracmm12$,
or to $\fracmm32{\pm}1$.
\vglue.1in

{\bf 4} {\it The string measure}.
Our starting point is the formal expression for scattering amplitudes,
$$
A_n\ =\ \sum_{h,k}\fracmm{1}{\cn}\int \! D(X\j g\c A)\,
e^{-S_{\rm m}}\,V_1\ldots V_n \quad,
\eqno(31)$$
where $S_{\rm m}$ is the gauge-invariant $N{=}2$ string (matter) action,
$V_\ell$ represent vertex operators for particles, and $\cn$ denotes the
volume of the gauge group.
We fix the $N{=}2$ superconformal gauge and use the BRST method \cite{bkl}.
A careful treatment of the Faddeev-Popov determinant, naively~\footnote{
Our notation is as follows (see refs.~\cite{klp,bkl} for more details):
$(b,c)$ stand for conformal ghosts,
$(\b^{\pm},\g^{\pm})$ for $N{=}2$ superconformal ghosts, and
$(\tilde{b},\tilde{c})$ for $U(1)$ ghosts.
The ghosts for the Weyl and super-Weyl symmetries are ignored
since they do not propagate. $S_{\rm gh}$ is the ghost action.}
$$
\int \! D(bc\b\g\tilde b\tilde c)\, e^{-S_{\rm gh}} \quad,
\eqno(32)$$
yields anti-ghost zero mode insertions for each moduli direction.
As we already know from $N{=}0$ and $N{=}1$ string theory \cite{dp},
these anti-ghost insertions come paired with the corresponding
Beltrami differentials which are the tangents to the moduli slice.
Let us take $h{>}1$ for simplicity; the cases of the sphere and torus
require obvious minor modifications due to isometries. We get
$$
\abs{\prod^{3(h-1)+n}_{m=1}\int_{\S}\m_m b\,}^2\ =\
\abs{\prod^{3(h-1)+n}_{m=1}\oint_{C_m}\! b\,}^2
\eqno(33)$$
for the metric moduli, where the Beltrami differentials
$(\m_m)^z_{\bar{z}}=\de_{\bar{z}}(v_m)^z$ have been represented in terms of
quasiconformal vector fields $v_m$ with a unit jump
across closed contours~$C_m\,$.
The commuting superconformal ghosts yield
$$
\abs{\prod_{a^+=1}^{2(h-1)+k+n}\d\Bigl(\b^+(z_{a^+})\Bigr)\quad
\prod_{a^-=1}^{2(h-1)-k+n}\d\Bigl(\b^-(z_{a^-})\Bigr)\,}^2
\eqno(34)$$
for the $N{=}2$ fermionic moduli, with a delta-function choice for the
fermionic Beltrami differentials. \footnote{
Like in the $N{=}1$ string, this raises issues of globality and
boundary terms in moduli space. We have no comment here.}
Finally, as a novel feature we obtain
$$
\prod_{i=1}^{h}\;\biggl[\oint_{a_i}\tilde{b}\;\oint_{b_i}\tilde{b}\biggr]
\quad \abs{ \prod_{\ell=1}^{n-1} \oint_{z_\ell}\tilde{b} \,}^2
\eqno(35)$$
for the $U(1)$ moduli, after taking the real abelian one-forms for the
$U(1)$ Beltrami differentials.
It must be added that this counting presumes that the vertex operators~$V_\ell$
are taken from the natural picture- and ghost-number sector, namely
$(\p^+,\p^-)=(-1,-1)$ and of $\tilde c c$ type. \footnote{
This allows only NS states; R states will be generated below.}

The $N{=}2$ supergravity fields enter the full (BRST-invariant)
$N{=}2$ string action $S_{\rm tot}=S_{\rm m}+S_{\rm gh}$
as Lagrange multipliers \cite{klp,bkl}.
Since the action $S_{\rm tot}$ is {\it linear\/}
in the fermionic and $U(1)$ moduli, we may integrate those out
and arrive at an additional insertion of
$$
\abs{
\prod_{a^+=1}^{2(h-1)+k+n} \! G^+_{\rm tot}(z_{a^+}) \,
\prod_{a^-=1}^{2(h-1)-k+n} \! G^-_{\rm tot}(z_{a^-}) \quad
\prod_{\ell=1}^{n-1} \d\Bigl( \oint_{z_\ell}\! J_{\rm tot} \Bigr)\,}^2 \quad
\prod_{i=1}^{h} \biggl[
\d\Bigl( \oint_{a_i}\! J_{\rm tot} \Bigr) \,
\d\Bigl( \oint_{b_i}\! J_{\rm tot} \Bigr) \biggr]
\eqno(36)$$
where $G^{\pm}_{\rm tot}$ and $J_{\rm tot}$ are the full (BRST-invariant)
supercurrents and $U(1)$ current of the $N{=}2$ string,
respectively \cite{klp,bkl}.

Combining eqs. (33)--(36), we find a product of
$N{=}2$ picture-changing operators ({\it cf.\/} ref.~\cite{pco}),
$$
\abs{
\prod_{m=1}^{3(h-1)+n}\!\oint_{C_m}\!b \;
\prod_{a^+=1}^{2(h-1)+k+n}\! Z^+(z_{a^+}) \,
\prod_{a^-=1}^{2(h-1)-k+n}\! Z^-(z_{a^-}) \;
\prod_{\ell=1}^{n-1} Z^0(z_\ell) \,}^2 \;
\prod_{i=1}^{h} \Bigl[ Z^0(a_i)\,Z^0(b_i) \Bigr] \,
\eqno(37)$$
where
$$
Z^{\pm}\ :=\ \d(\b^{\pm})\,G^{\pm}_{\rm tot}\ =\ \{Q_{\rm BRST},\x^{\pm}\}
\eqno(38)$$
and
$$
Z^0(\g)\ :=\ \Bigl( \oint_{\g} \tilde{b} \Bigr) \;
\d \Bigl( \oint_{\g} J_{\rm tot} \Bigr)
\eqno(39)$$
for a closed contour~$\g$ (being a homology cycle or encircling a puncture).
Up to $n$ of the $b$ insertions can be used to convert $c$-type
vertex operators to integrated ones, and any number of $Z^\pm$ may be taken
to upwardly change their $(-1,-1)$ pictures, keeping the total picture
at $(2(h{-}1){+}k,2(h{-}1){-}k)$.
The novel $U(1)$ picture-changing operators $Z^0$ serve two purposes:
First, a $Z^0$ associated with a homology cycle enforces a projection onto
charge-neutral excitations propagating across the loop and ensures
factorization on neutral states when pinching the cycle.
Second, a $Z^0$ attached to a puncture can be absorbed by the
$\tilde c$-type vertex operator,
changing its $U(1)$ ghost number by removing the $\tilde c$ factor.
In this way, only a single $\tilde c$-type vertex operator
remains in the end, plus the $2h$ real $Z^0$ insertions associated with
the homology cycles.
Thus not only $Z^\pm$ but also $Z^0$ provides a map between BRST cohomology
classes. \footnote{
Neither $Z^\pm$ nor $Z^0$ have a local inverse \cite{bkl},
while $Z^0$ is nilpotent.}

Eq.~(37) is formally BRST-invariant, which is important for the consistency and
BRST-invariance of $N{=}2$ string amplitudes. The integration over the $N{=}2$
matter fields does not present a principal problem, but it has to be done in
the presence of a background instanton field $A^{\rm inst}$ minimally coupled
to the fermions.
No sum over fermionic spin structures appears since it has already been
carried out as part of the $U(1)$ moduli integration. The integration over
the metric moduli follows the lines of the familiar $N{=}0$ and $N{=}1$ cases.

{\bf 5} {\it The spectral flow}.
The $N{=}2$ NSR string fermionic matter fields $\j^{\pm\m}$ are sections of
a complex twisted spinor bundle, just like the gravitini.
For vanishing instanton number, $k{=}0$, this becomes a twisted holomorphic
spinor bundle.  The associated spin structures parametrizing the NS/R sectors
or monodromies of $\j$ in an $n$-point function are labeled by the half-points
$(\fracmm12{\bf Z}/{\bf Z})^{2(h+n-1)}$ in the Picard variety of eq.~(16).
Now observe that by a unitary transformation via
$$
U(z)\ =\ \exp\Bigl\{ i\int^z_{z_0} \hat{A} \Bigr\}
\eqno(40)$$
of a given spinor bundle we can always change the monodromies with a suitable
$\hat{A}\in\{A^{\rm Teich}\}$, because the $\j$ carry $U(1)$ charge \cite{amv}.
We can in fact reach any point in the Picard variety and, in
particular, move to any other spin structure.
Since the unitary transformation of eq.~(40) is equivalent to a shift in the
integration variables $A^{\rm Teich}\to A^{\rm Teich}+\hat{A}$
and the $U(1)$ moduli space has no boundary,
we conclude that the sum over the NSR spin structures
(and, in fact, over all intermediate monodromies)
is automatically contained in the integration over the $U(1)$ moduli.
This is the so-called {\it spectral flow\/} of Ooguri and Vafa \cite{ov}.

As a consequence, the distinction between NS and R sectors is
$U(1)$ moduli-dependent and thus cannot be physical.  This feature is not
restricted to $h\geq 1$, but appears just as well for the $n$-punctured
sphere, i.e. for tree-level $N{=}2$ string amplitudes. More precisely, any
pair of R-type punctures can be turned into a pair of NS-type,
since there are $n{-}1$ independent cycles and the sum of all twists
has to vanish.
Hence, the R-type and NS-type states of the $N{=}2$ string cannot be
physically distinguished and their correlators must coincide, which is
consistent with our previous explicit calculations \cite{bkl}.
Note, however, that the correlation functions should depend on the value~$k$
of~$c$, which may be changed by allowing for a non-zero but integral total
twist of the external states.

A shift in the $U(1)$ puncture moduli, $\l^+_\ell\to\l^+_\ell+\q_\ell$,
which shifts the puncture monodromies by $\q_\ell$,
also modifies the vertex operators~$V_\ell(z_\ell)$ present in
the string path integral,
$$
V_\ell(z_\ell)\ \longrightarrow\
V_\ell^{\q_\ell}(z_\ell)\ \equiv\ \sfo(\q_\ell,z_\ell)\,V_\ell(z_\ell)\quad,
\eqno(41)$$
where, remarkably, the spectral-flow operator $\sfo(\q,z)$ can be written
in the explicit form \footnote{
Normal ordering is implicit in our formulae.
$\sfo(\q)$ generalizes the instanton number-changing operators $\sfo(\pm1)$
of refs.~\cite{bv2,b2}.}
$$
\sfo(\q,z)\ =\ \exp\Bigl\{ \q\int^z \! J_{\rm tot}(z')dz' \Bigr\}\
=\ \exp\bigl\{ \q\,(\f^+ -\f^- -\vf^+ +\vf^- +\tilde{b}c)(z) \bigr\}~.
\eqno(42)$$
Here, we have bosonized \cite{bkl}
$$
J_{\rm tot}\ =\ \{Q_{\rm BRST},\tilde{b}\}\
=\ -\fracmm12\j^+{\cdot}\j^- +\pa(\tilde{b}c)+\fracmm12(\b^+\g^- {-}\b^-\g^+)
\eqno(43)$$
in the holomorphic basis via ($\e{=}\pm1$)
$$
\j^{\pm,\e} = 2e^{\e\f^{\pm\e}} \quad,\qquad
\g^\pm = \h^{\pm}e^{+\vf^{\pm}} \quad,\qquad
\b^\pm = e^{-\vf^{\mp}}\pa\x^{\pm} \quad,
\eqno(44)$$
and obtained a {\it local\/} operator.
It is easy to check that $\sfo$ is BRST invariant but not BRST trivial.
However, like for the $Z^\pm$, the derivative $\pa\sfo$ is BRST trivial,
so we may move spectral-flow operators around at will on the world-sheet.
Therefore, an insertion of $\prod_\ell \sfo(\q_\ell,z_\ell)$ is
BRST-equivalent to unity as long as $\sum_\ell \q_\ell=0$, implying again
the invariance of string amplitudes under shifts of the puncture monodromies,
i.e.
$$
\Bigl\langle V_1 V_2\ldots V_n \Bigr\rangle\ =\
\VEV{V_1^{\q_1} V_2^{\q_2}\ldots V_n^{\q_n}} \qquad{\rm for}\qquad
\sum_\ell \q_\ell =0 \quad.
\eqno(45)$$

The $N{=}2$ superconformal algebra generated by
$(T_{\rm tot}, G^\pm_{\rm tot}, J_{\rm tot})$
is extended to the {\it small\/} $N{=}4$ superconformal algebra by
adding the $SU(2)$ ladder operators $\sfo(\pm1)$ and closing the algebra.

The NS$\leftrightarrow$R exchange is accomplished at, $\q=\pm\fracmm12$,
which may be symbolically written as $\sfo^\pm{\rm NS}={\rm R}^\pm$ where
$\sfo^\pm\equiv\sfo(\pm\frac12)$ and ${\rm NS(R)}\equiv V_{\rm NS(R)}$.
The index on R indicates that we can flow to two different Ramond vertex
operators.
The $N{=}2$ string vertex operators are known to exist in different
(holomorphic) pictures $(\p^+,\p^-)$,
which are connected by the process of picture-changing
$(\p^+,\p^-)\to(\p^+{+}1,\p^-)$ or $(\p^+,\p^-)\to(\p^+,\p^-{+}1)$
via the picture-changing operators $Z^+$ or $Z^-$ of eq.~(38), respectively.
The spectral flow operators $\sfo^{\pm}$ can actually be interpreted as
yet additional picture-changing operators:
$$
\sfo^\pm:\qquad (\p^+,\p^-)\ \to\ (\p^+{\pm}\frac12,\p^-{\mp}\frac12)\quad.
\eqno(46)$$
This leads, for example, to the following identifications among tree-level
two-, three- and four-point functions at instanton number $k{=}0$:
$$\eqalign{
&\Bigl\langle{\rm NS}\;{\rm NS}\Bigr\rangle\ =\
\VEV{{\rm R}^+\,{\rm R}^-}\quad,\quad
\Bigl\langle{\rm NS}\;{\rm NS}\;{\rm NS}\Bigr\rangle\ =\
\VEV{{\rm NS}\;{\rm R}^+\,{\rm R}^-} \quad,\cr
&\Bigl\langle{\rm NS}\;{\rm NS}\;{\rm NS}\;{\rm NS}\Bigr\rangle\ =\
\VEV{{\rm NS}\;{\rm NS}\;{\rm R}^+\,{\rm R}^-}\ =\
\VEV{{\rm R}^+\,{\rm R}^-\,{\rm R}^+\,{\rm R}^-}\ =\ 0
\quad,\cr}
\eqno(47)$$
which were all verified by explicit calculations in ref.~\cite{bkl}.
At tree-level, non-vanishing correlators require $\abs{k}\leq n{-}2$,
so that the complete three-point amplitude, for example,
also has $k{=}\pm1$ contributions.
These can be generated from $k{=}0$ by inserting
$\sfo(\pm1)=\sfo^\pm\sfo^\pm$ into $\VEV{{\rm NS}\,{\rm NS}\,{\rm NS}}$,
resulting in
$$
\VEV{{\rm NS}\;{\rm R}^+\,{\rm R}^+}\ +\ \VEV{{\rm NS}\;{\rm R}^-\,{\rm R}^-}
\quad.
\eqno(48)$$
Noting that $V_{\rm R}^-=f(p)V_{\rm R}^+$ with a momentum-dependent factor
$f(p)$ \cite{bkl}, we relate eq.~(48) to the $k{=}0$ case and find that the
two terms in eq.~(48) cancel each other on-shell.
This leaves us with the standard $U(1,1)$-invariant result for the tree-level
three-point function and removes the $U(1,1)$ non-invariant $k{\neq}0$ terms
(see also ref. \cite{lp}).
It would be interesting to check whether such a mechanism ensures global
$U(1,1)$ symmetry in general.

Finally, it should be noticed that the discussion of spectral flow
heavily relied on the use of the {\it holomorphic\/} basis for
bosonization. In contrast, in the {\it real\/} basis (discussed also at length
in refs.~\cite{bkl,lp}) the spectral flow is obscured and the operator
defined by eq.~(43) is non-local.

With pleasure we thank the participants of the conference ``Strings'95''
at the University of Southern California in March 13--18 and, in particular,
Nathan Berkovits, Hong L\"u, Hiroshi Ooguri and Chris Pope,
for fruitful discussions.
Our special gratitude to the crew of the British Airways flight \# 268,
from Los Angeles to London, where this work was completed.
\vglue.2in

\end{document}